# COMPARISON OF ITS-90 REALIZATIONS FROM 13 K TO 273 K BETWEEN LNE-CNAM AND INRIM

**D Imbraguglio[1), 3)], P P M Steur[1)], F Sparasci[2), 3)]**
*1) Istituto Nazionale di Ricerca Metrologica, Torino, Italy*
*2) LNE-CNAM, La Plaine Saint-Denis, France*
*3) Joint Research Laboratory for Fluid Metrology Evangelista Torricelli between LNE and INRIM*

*E-mail (corresponding author): d.imbraguglio@inrim.it*

**Abstract**

Since Key Comparison CCT-K2, only few comparisons as regards realizations of cryogenic fixed points have been carried out, either bilateral (three, of which one still pending) or multilateral (two trilateral). Since a forthcoming follow-up is unlikely, any bilateral comparison, apart from key comparisons, is most welcome to provide evidence of the continuing thermometric capabilities of the laboratories involved in the realization of the ITS-90.

Not too long after the establishment in 2018 of the Joint Research Laboratory for Fluid Metrology "Evangelista Torricelli" between LNE-CNAM and INRIM, the two laboratories agreed to perform a bilateral comparison, involving all the cryogenic fixed points of the scale between 13.8033 K and 273.16 K - including the two hydrogen vapour-pressure temperatures -, required to cover the full platinum resistance-thermometer range, as dealt with in section 3.3.1 of the text underlying the ITS-90. The results are reported here and show a substantial agreement within the combined uncertainties.

## 1 Introduction

Both INRIM and LNE-CNAM were among the laboratories participating in the Key Comparison 2 [1], promulgated in the 90's by the Consultative Committee for Thermometry (CCT-K2) to address the realizations of the International Temperature Scale of 1990 (ITS-90) from 13.8033 K to 273.16 K [2]. Since then, only a limited number of Key Comparisons have been conducted among national measurement institutes (NMIs), only bilateral or trilateral ones [3-6], and a full-fledged international effort for a repeat of CCT-K2 seems highly unlikely in the near future. Some have sustained that one can recognize the not too recent (1997-2005) international comparison of fixed point cells performed at PTB [7] as such but, since the measurements were done exclusively at the pilot's laboratory, no information can be extracted about the capabilities of the other participating laboratories, exactly the reason why Key Comparisons are performed. Therefore, the present bilateral comparison between LNE-CNAM and INRIM, under the vestiges of the recently created Joint Research Laboratory for Fluid Metrology "Evangelista Torricelli", is considered a most welcome addition to the still scarce verification of the ongoing maintenance of activities concerning the realization of the ITS-90 in the cryogenic range.

ITS-90 is defined by standard platinum resistance thermometers (SPRTs) from 13.8033 K up to 1234.93 K. To cover the whole cryogenic range between 13.8033 K and 273.16 K, dealt with in section 3.3.1 of reference [2], capsule-type SPRTs (CSPRTs) are usually employed. CSPRTs are calibrated at eight defining fixed points of the scale, of which six triple points (TPs) and two vapor-pressure points (VPPs). Both LNE-CNAM and INRIM are equipped with cells for the six TPs, namely those of equilibrium hydrogen (e-$H_2$, $T_{90}$ = 13.8033 K), neon (Ne, $T_{90}$ = 24.5561 K), oxygen ($O_2$, $T_{90}$ = 54.3584 K), argon (Ar, $T_{90}$ = 83.8058 K), mercury (Hg, $T_{90}$ = 234.3156 K) and water (TPW, $T_{90}$ = 276.16 K), see Table 1.

The first four TPs are realized in similar facilities, basically two twin isothermal cryostats [8, 9] provided with multiple TP cells. These systems allow measurements to be performed, in a single temperature excursion run, employing the calorimetric method, as recommended in [10] for low-temperature fixed points. The method consists in heating the sample through the phase transition under nearly isothermal conditions, by an intermittent input of heat. After each heat pulse, the cell is allowed to come to thermal equilibrium. Details are provided in section 2.1.1 (the reader is also referred to the Appendix of the present paper for details about the overall calibration procedure). The Hg TP is realized in systems which differ widely in the two laboratories. INRIM employs cells immersed in a bath supplying continuous heating [11], while LNE-CNAM uses miniature cells in an isothermal cryostat [12], and realizes the phase transition on the basis of the calorimetric method [10]. The TPW is realized in the same way, with quartz sealed cells inside a temperature-controlled bath [13]. Finally, in both laboratories, the two VPPs of e-$H_2$, near 17.0 K and 20.3 K, see Table 1, are not systematically realized with open cells or a gas thermometer: they are rather transferred with rhodium-iron resistance thermometers (RIRTs) previously calibrated against an *interpolating constant-volume gas thermometer* (ICVGT), either by the same lab (INRIM) or by another lab – see sections 2.1.2 and 2.2.2.

The comparison described in this paper involves the realization of the ITS-90 between the TP of e-$H_2$ and the TPW (see Table 1). It was carried out by the exchange of a CSPRT (Rosemount 162D, serial number (s/n): 5435) as a transfer standard between the two NMIs, calibrated at LNE-CNAM first, then at INRIM. The comparison not being officially registered in the CIPM[1] or EURAMET[2] databases, a specific protocol has not been defined. Therefore, to ensure measurement comparability, the two laboratories have applied the procedures registered within their respective quality management systems, regularly employed to produce calibration certificates, compliant with the *Calibration and measurement capabilities* (CMCs) included in the CIPM *Mutual Recognition Arrangement* (MRA) and available in the *Key Comparison Database* (KCDB) [14].

## 2 Experimental Apparatus and Procedures

### 2.1 Calibration at INRIM

#### 2.1.1 Triple point cells and direct-calibration apparatus below 84 K

The TP cells constituting the *Italian National Temperature Standard* below 84 K are given with the serial numbers Ec1H2 for the TP of e-$H_2$, Ec2Ne for the TP of Ne, Ec1O2 for the TP of $O_2$ and Ec1Ar for the TP of Ar. The designation Ec refers to element (E) of type c, as developed during the so-called MULTICELLS project [15, 16], the four cells being able to be mounted together on a copper block, allowing a further five thermometers to be put in direct thermal contact. With respect to the e-$H_2$ and Ne cells, information is required regarding their isotopic composition: cell Ec1H2 contains 91.6 µmol D / mol H [17] while cell Ec2Ne contains 0.093 36 mol $^{22}$Ne / mol Ne and 0.002 649 mol $^{21}$Ne / mol Ne [18]. As regards the calibration of CSPRT-5435 (Transfer standard of the present work) the isotopic corrections, inherent in the reference points, were applied as prescribed in the *Technical Annex of the Mise-en-Pratique of the kelvin*

---

[1] CIPM: International Committee for Weights and Measures

[2] EURAMET: European Association of National Metrology Institutes

[19], now after the redefinition of the kelvin, denominated *Technical Annex for the International Temperature Scale of 1990* (ITS-90) [20]. The values of the respective corrections for the isotopic effect, in kelvin, were found to be -14 μK for e-$H_2$ and -126 μK for Ne.

As outcome, the MULTICELLS project produced also an automated data-acquisition system [8] with the software requiring always a control thermometer to set the temperature of the copper block. Since its inception this is a RIRT, designated B191. The other (capsule) thermometers are usually CSPRTs. Thus, the computer-controlled system for direct calibration at the TPs consists of the copper block, housing the four cells and the capsule thermometers (SPRT / RIRT), surrounded by an adiabatic copper shield, whose temperature is further controlled by a RIRT (s/n 10729) and a heater. The role of the adiabatic copper shield is to create the closest approximation of the adiabatic state for the cell. It is enclosed, on its turn, by a shield connected to the second stage of a Gifford-McMahon cryocooler [8]. This shield is temperature-controlled only for the TPs of $O_2$ and Ar in order to supply adequate thermal conditions for the adiabatic shield. Before realizing a TP, the system performs automated checks on thermal capacitances and on the degree of adiabaticity by means of the thermometer B191. Then, a proper plateau realization is started: a series of intermittent heat pulses is applied to the solid TP substance, realizing a series of states of increasing melted fraction ($F$), see in the Appendix, until the liquid state has been reached. After each heat pulse, when the cell has returned to thermal equilibrium again, the resistances for all the thermometers are measured in sequence using a DC resistance bridge (Measurements International, model 6010T). Depending on the temperature of the TP, the measurement current (1 mA, 2 mA or 5 mA) and the reference resistor (10 Ω or 1 Ω, temperature-controlled in a bath) are chosen in order to maximize the sensitivity of the CSPRT and that of the resistance bridge at the given temperature. The melted fractions $F$ after each heat pulse are calculated taking account of the power applied via the pulse and its duration, the residual heat flux due to the non-ideal adiabaticity and the additional heat generated by the measuring current of the thermometers. Finally, the measured resistance values (reduced to zero current) of the CSPRT are plotted against the values $1/F$ and extrapolated to $F = 1$, see Figure A.4 in the Appendix.

2.1.2 Vapor-pressure points of e-$H_2$ and comparison-calibration apparatus between 4 K and 27 K

Following the prescriptions of the ITS-90, a CSPRT calibration for the range dealt with in section 3.3.1 of reference [2], the full range from 13.8033 K to 273.16 for CSPRTs at cryogenic temperatures, needs two additional calibration points compared to those required by the three intermediate sub-ranges 3.3.1.1, 3.3.1.2 and 3.3.1.3: one close to 17.0 K and another close to 20.3 K. Explicitly citing ITS-90: "*These last two may* *be determined either: by using a gas thermometer as described in Sect. 3.2, in which case the two temperatures must lie within the ranges 16.9 K to 17.1 K and 20.2 K to 20.4 K, respectively; or by using the vapour pressure-temperature relation of equilibrium hydrogen (omitted)*". In the past, IMGC (now INRIM) has realized the ICVGT (see Introduction) as prescribed by the ITS-90 [21-23]. The INRIM RIRT labelled 232324 carries this ICVGT scale and the International Comparison Euramet.T-K1 [24] corroborates the results. This scale was transferred in recent years to as many RIRTs as possible, in order to reduce the risk of losing the calibration of this single thermometer, RIRT 232324. Besides, thanks to a recent Euramet approval [14] of a CMC regarding the calibration by comparison of CSPRTs in the 4 K – 27 K range, the two VPPs can now be transferred directly via the RIRT 232324, thus allowing easier comparisons between different realizations of the scale even down to 4 K.

In order to disseminate all out such a historical reference, RIRT 232324 is mounted on another devoted cryostat, cooled by a pulse-tube cryocooler (Cryomech, model PT410) which minimizes mechanical vibrations. Here the copper block, housing both the RIRT and the thermometers to be compared (CSPRTs or RIRTs), is enclosed by an adiabatic shield performing the temperature control directly via the 232324 readings. Measurements at the VPP temperatures are still performed via an automated software which pilots an AC resistance bridge (Automatic Systems Laboratories, model F900) connected to a temperature-controlled 10 Ω reference resistor. Thermometer currents are selected such that the powers generated by the mounted thermometers differ as little as possible, and thus with little perturbation of the thermal condition of the comparison block during switching.

### 2.1.3 Overall calibration procedure of the transfer standard

The transfer standard underlying the comparison, CSPRT 5435, was transferred from LNE-CNAM to INRIM without any information about previous calibrations or traceable applications, except for its resistance value at the TPW. The calibration procedure adopted at INRIM was the following: firstly the transfer standard was calibrated at the TPW and at the TP of Hg using a home-made stem-housing conduit, and provided with a copper / stainless steel jacket, to avoid liquid infiltration. In the case of the TPW, the *Italian National Temperature Standard*, used in this stage, was the cell Hart model 5901, s/n 1322 (Hart Scientific, part of Fluke Corp.), while for the TP of Hg, the home-made cell CRYO-07 was used. The corrections applied to the temperature values due to the depth immersion within the cells were 0.163 mK and -0.891 mK, respectively, while a further correction of 0.077 mK was added to the temperature values measured at the TPW to account for the isotopic composition, as established in [20]; for details see, in particular Table 2, as regards uncertainties correlated with INRIM results and, in general, Section 4.

The measurement at the TPW showed a difference of only -0.057 mK with respect to the value supplied by LNE-CNAM. As the thermometer remained stable after transport, it was then mounted in the cryostat for direct calibration over the temperature range from 13.8033 K up to 83.8058 K, and the whole system was cooled to its minimum (≈ 10 K) using He as exchange gas. The gas was pumped out overnight, and a temperature of about 14 K (only about 0.2 K above the TP of e-$H_2$) was set for the adiabatic shield in order to allow for the ortho-para conversion [10] to take place as fast as possible (see the Appendix). Then, the system was cooled down to about 0.5 K below the TP of e-$H_2$ (13.8033 K), and the fully automatic plateau realization was started, taking about 3-4 days to complete. In this way, three plateaus were realized for the TP of e-$H_2$ to confirm the reproducibility and, similarly, three plateaus for each of the other cryogenic TPs (Ne, $O_2$ and Ar) at 24.5561 K, 54.3584 K and 83.8058 K. Reproducibilities, derived from these realizations, were 0.021 mK, 0.027 mK, 0.037 mK and 0.115 mK, for the TPs of e-$H_2$, Ne, $O_2$ and Ar, respectively.

After completion of the third plateau for Ar, the system was allowed to return to room temperature, and the transfer standard (CSPRT 5435) was mounted on the pulse-tube system for the comparison measurements with RIRT 232324, carrying the two VPPs of the ICVGT scale, see the Introduction. The temperatures near 17 K and 20.3 K were regulated by a temperature controller after the system was cooled down to its minimum base temperature (≈ 4 K), the overall measurement process taking about two weeks to complete. INRIM performed VPP measurements at $T_{17K,INRIM} = 17.035$ K and $T_{20K,INRIM} = 20.27$ K, while LNE-CNAM measured VPPs at $T_{17K,LNE\text{-}CNAM} = 17.0332$ K and $T_{20K,LNE\text{-}CNAM} = 20.2687$ K (see sections 2.2.2 and 2.2.4). All these temperature values fall inside the ranges identified in the ITS-90 [2] and enable valid realizations of the ITS-90 itself. However, to allow a direct comparison between INRIM and LNE-CNAM results, INRIM had to scale its measured values, to produce computed calibration values at $T_{17K,LNE\text{-}CNAM}$ and $T_{20K,LNE\text{-}CNAM}$. In particular, two quantities $\Delta R_{17K} = (T_{17K,INRIM} - T_{17K,LNE\text{-}CNAM})\cdot dR/dT_{17K}$ and $\Delta R_{20K} = (T_{20K,INRIM} - T_{20K,LNE\text{-}CNAM})\cdot dR/dT_{20K}$ were subtracted respectively from the resistances measured by INRIM at $T_{17K,INRIM}$ and $T_{20K,INRIM}$, with $dR/dT_{17K}$ and $dR/dT_{20K}$ being the sensitivity coefficients of the CSPRT. As for the measurements at the other fixed points, the self-heating of the CSPRT was always taken into account by measuring each time the thermometer resistance at $x$ mA and $x\cdot\sqrt{2}$, then, by extrapolation to zero current. Finally, after the comparison, the transfer standard was calibrated one last time at the TPW by means of the cell Hart model 5901C-Q, s/n1015 (Hart Scientific), as an internal stability check, resulting in no significant shift.

## 2.2 Calibration at LNE-CNAM

### 2.2.1 Triple point cells and calibration apparatus below 84 K

The TP cells used at LNE-CNAM below 84 K are of a different design but derive, like the INRIM ones, from the same MULTICELLS project described in section 2.1.1. They are the same used in the trilateral comparison [5], identified with the following serial numbers: $H_202/1$ for e-$H_2$, Ne02/1 for Ne, $O_201/1$ for $O_2$, Ar01/1 for Ar. They are assembled to form a solid block where up to four thermometers can be installed. An additional copper block can be mounted under the cell assembly, to allow the fitting of four extra thermometers. No information on isotopic composition is available on e-$H_2$ and Ne cells, thus an extra uncertainty on isotopic composition is included in the calibration uncertainty budget (see tables B2 and B4 in [5]).

A cryostat, based on the Gifford-McMahon cryocooler [9], similar to that available at INRIM (section 2.1.1), is employed at LNE-CNAM. The cell assembly with the thermometers is placed inside an adiabatic copper shield, whose temperature is controlled by means of a Cernox® thermometer (Lakeshore) and an associated heater. The adiabatic shield is surrounded by a shield connected to the second stage of the cryocooler, whose temperature is controlled with another Cernox® thermometer and an associated heater. The phase transitions are realized by means of a series of intermittent heat pulses and the thermometers under calibration are measured with an AC resistance bridge (Automatic Systems Laboratories, model F18). As for INRIM's system, probe currents and temperature-controlled reference resistors are selected to maximize the sensitivity.

### 2.2.2 Vapor-pressure points of e-$H_2$

LNE-CNAM usually realizes calibration of thermometers close to 17.0 K and 20.3 K by comparison with one or more reference thermometers. In this case the temperatures were 17.0332 K and 20.2687 K. The batch of reference thermometers consists of four RIRTs and three CSPRTs calibrated by other NMIs and three additional RIRTs calibrated at LNE-CNAM by comparison with the former. For this comparison, RIRT 229071 and CSPRT B398 were used as reference thermometers. The first was calibrated at LNE-CNAM by comparison with RIRTs B115 (NIST calibration) and B188 (NPL calibration), while the second was calibrated at NPL.

Reference thermometers are installed in the same cell assembly as the thermometers under calibration and the measurements are performed with the same cryocooler and at the same temperature excursion realized for the TPs measurements. In this way, the reference thermometers are measured also at the TP temperatures, allowing an in-situ check of the stability of their calibration.

### 2.2.3 Triple point of Hg and associated apparatus

In 2013, LNE-CNAM developed a calorimeter for the simultaneous calibration of long-stem and CSPRTs at the TP of Hg [12]. The difference with most traditional systems [11] is the capability to perform calorimetric realizations of the phase transition, as the apparatus is essentially an isothermal cryostat. Capsule thermometers are installed in a copper block connected to the TP cell and the assembly is surrounded by a thermal shield. The latter is thermally stabilized by another mercury triple point cell mounted above the former, here named "guard cell", which plays a role in sinking the heat flow conveyed by the well of the long-stem SPRT. Heaters are installed on both the TP and the guard cells and the second is also equipped with a control thermometer. The whole system is contained in a vacuum can and immersed in a temperature-controlled bath. Mercury is preliminarily frozen in both cells, then the bath temperature is slightly raised above the TP temperature of Hg (typically by about 10 mK) and the TP of Hg in the guard cell is induced by applying a series of heat pulses through its heater. Heating of the guard cell is stopped when the fraction of melted Hg is around 30% of the total: at this point, the ongoing phase transition in the guard cell stabilizes the shield temperature (reducing temperature oscillations to less than 1 mK) and sinks any heat arriving from the piping and the central well housing the long-stem SPRT. Afterwards, heat pulses are applied to the main TP cell: the phase transition is realized in the same way as for cryogenic TPs; the melting plateau lasts generally 4 to 5 days. The Hg apparatus of LNE-CNAM is shown in Figure 1.

### 2.2.4 The transfer standard: calibration procedure

As mentioned in section 2.1.3, the transfer standard CSPRT 5435 was calibrated at LNE-CNAM before sending it to INRIM. All the TP cells used were part of the *French National Temperature Standard* batch. First, the thermometer was calibrated at the TPW using a home-made adaptor stem, into the cell Hart model 5901, s/n 1422 (Hart Scientific). Then, the thermometer was calibrated in the TP cell of Hg, s/n 08144 (home-made, described in 2.2.3). After this step, the stability of the thermometer was checked at the TPW and no significant shift was observed with respect to previous measurements. In all these measurements, self-heating was evaluated by means of the same procedure as followed by INRIM, i.e. the thermometer resistance was extrapolated to zero current from measurements performed at 1 mA and 1·√2 mA.

Later on, CSPRT 5435 was installed in the low-temperature cryostat for fixed points, inside a cell assembly composed by the cells $H_2$02/1, Ne02/1, $O_2$01/1 and Ar01/1, referred to earlier in section 2.1.1. The cryostat was cooled down to the base temperature (below 10 K) with exchange gas, then the gas was removed, and the temperature was set to 14 K for three weeks, to allow the ortho-para conversion of $H_2$. Then, the system was cooled down to about 0.5 K below the TP temperature of e-$H_2$ to freeze the sample and an automatic plateau realization was initiated, taking about 4 days to complete. Self-heating was evaluated on each thermometer, from measurements performed at 5 mA and 5·√2 mA on CSPRTs, and 1 mA and 1·√2 mA on RIRTs. All the thermometers installed in the cell assembly were cyclically measured with a scanner, including the transfer standard CSPRT 5435 and the two reference thermometers RIRT 229071 and CSPRT B398. The e-$H_2$ plateau was repeated three times. A similar measurement procedure was employed at higher temperatures, to realize the TPs of Ne, $O_2$ and Ar. At the TPs of $O_2$ and Ar, currents of 2 mA and 2·√2 mA were supplied on CSPRTs, and 1 mA and 1·√2 mA on RIRTs.

After the e-$H_2$ realization, the temperature of the cell assembly was raised to 17.0332 K. RIRT 229071 was used as reference sensor, to control the temperature of the cell assembly. It was disconnected from the main resistance bridge, exclusively connected to an additional AC resistance bridge and powered with 1 mA, in order to be continuously measured and allow the control of the assembly temperature. The second reference thermometer, CSPRT B398, was kept connected to the main resistance bridge, and so it was cyclically measured with the scanner, and compared with the RIRT 229071. CSPRTs were all measured using 5 mA, but self-heating evaluations were not performed, as the additional heat produced by the thermometers when they were powered with 5·√2 mA would have significantly changed the temperature setpoint of the cell assembly. Thus, self-heating was inferred from measurements at the e-$H_2$ and Ne TPs, with a standard uncertainty of around 0.5 mK. The comparison at 17.0332 K was repeated three times, then the cryostat temperature was raised to 20.2687 K, applying the same procedure used at 17.0332 K.

Measurements were analyzed after the realization of the experiments. The correlated uncertainty values considered, dealing with LNE-CNAM only, can be found in Table 3 of Section 4. At this stage it was noticed that the resistance of the reference thermometer RIRT 229071 had shifted by -1,73(23) mΩ at the TPs of e-$H_2$ and Ne, with respect to its previous calibration done at LNE-CNAM. Such a resistance shift corresponded to -7.36 mK at 17.0332 K and -8.58 mK at 20.2687 K. Data from the second reference thermometer B398, which didn't show any instability at the TPs of e-$H_2$ and Ne, confirmed these shifts and the necessity to introduce corrections on the readings of RIRT 229071. Temperature corrections corresponding to the aforementioned shifts were applied to the calibration of the CSPRT 5435 at the two VPPs realized at LNE-CNAM. Standard uncertainties associated to these corrections were evaluated as 0.98 mK at 17.0332 K and 1.14 mK at 20.2687 mK.

After the realization of the fixed points below 84 K, the stability of the CSPRT 5435 was checked again at the TPW, and no significant difference was found with respect to previous measurements (less than 20 µK).

**3 Results**

All the procedures applied to carry out measurements and calculate comparison results are those regularly registered in the quality management systems of LNE-CNAM and INRIM, which ensure international traceability and compliance with the CIPM MRA, see Introduction. A consequence of this choice is that measurement uncertainties cannot be lower than those declared by each laboratory as CMCs, hereinafter referred to as $u_{MRA}$ (with a coverage factor $k = 1$, where the coverage factor refers to the confidence level, in this case, approximately 68%). However, they might be higher if the transfer standard shows instability or poor quality. To limit this risk, LNE-CNAM selected CSPRT 5435 in the batch of its most suitable CSPRTs: the choice revealed to be appropriate, as neither LNE-CNAM nor INRIM had to introduce any additional uncertainty to account for possible CSPRT quality lack. As shown in the following sections, the agreement between INRIM and LNE-CNAM is good at all fixed points between 13.8033 K and 273.16 K. In addition to the CMC uncertainties, INRIM and LNE-CNAM evaluated their "best" standard measurement uncertainties $u_{best}$ ($k = 1$) which, at some fixed points, might be lower than the CMC ones (see Section 4 for details). These results are not associated with the CMCs and may be considered as possible improved uncertainty budgets that may replace CMCs in the future.

Table 1 summarizes the overall results of the calibrations performed at INRIM and LNE-CNAM on the basis of the CSPRT 5435 transfer standard. $T_{90}$ values are the fixed point temperatures specified in the ITS-90 for the range 3.3.1, see Introduction, including the two VPP temperatures as defined in the scale. $W$ ratios are obtained by INRIM (labelled "i") and LNE-CNAM (labelled "j") according to equation (7) in the ITS-90 [2], i.e. $W(T_{90}) = R(T_{90}) / R(273.16\ \mathrm{K})$, where $R$ are resistance values corrected for all the effects mentioned in Section 2. Both $u_{MRA}$ and $u_{best}$ standard uncertainties are reported here directly in millikelvins with a coverage factor $k = 1$. $dT/dW$ are the mean sensitivity values of the thermometer calculated at each fixed point from the $W$ values, computed with the respective conversion software according to the deviation function (12) in [2], while $dT_{ij}$ are the temperature differences between the two laboratories, computed as $dT_{ij} = (dT/dW)\cdot(W_i - W_j)$. Lastly, the degree of equivalence, $d_{0.95}$, between the two NMIs is calculated on the basis of the expression [25]:

$$d_{0.95} \approx |dT_{ij}| + a \cdot \{1.645 + 0.3295 \cdot exp[-4.05(|dT_{ij}|/u_p)]\}u_p,$$

where

$$a = 0.283 + 0.717 \cdot 1.96 + 0.042 \cdot 1.96^3 \cdot exp\left[-0.399(|dT_{ij}|/u_p)^2\right],$$

assuming infinite degrees of freedom (a numerical approximation sufficient for all practical purposes), and $u_p$ is the combined standard uncertainty of the two laboratories, either obtained by combining $u_p$ with $u_{MRA}$ values ($u_{p,MRA}$), or $u_p$ with $u_{best}$ values ($u_{p,best}$), see Table 1. The results are displayed in Figure 2 and Figure 3 for $u_{p,MRA}$ and $u_{p,best}$, respectively. In both cases LNE-CNAM and INRIM measurements agree within the combined standard uncertainties for each fixed point in the range. When considering MRA uncertainties, the degree of equivalence, $d_{0.95}$, is comprised between 0.78 mK and 1.63 mK if direct calibrations at the TPs are performed, whereas it goes up to about 9 mK at the VPPs calibrations realized by comparison. The latter is found to decrease by about 3 mK if best uncertainties are considered instead, with the degree of equivalence at the TPs comprised between 0.39 mK and 1.53 mK. As somehow predictable, however, these results confirm qualitatively, in the end, the lower quality of calibrations achieved by comparisons with respect to the ones obtained by direct realizations of the fixed points. Comparable uncertainties can only be obtained using the vapour-pressure – temperature relation of hydrogen directly, requiring however a dedicated apparatus.

**4 Estimated Uncertainties**

Tables 2 and 3 list the uncertainty budgets of INRIM and LNE-CNAM, respectively, on the calibrations of CSPRTs in the range 3.3.1 of the ITS-90, see Introduction, with details given for each fixed point on the different uncertainty components. Only "best" uncertainty budgets $u_{best}$ are reported here, as information on $u_{MRA}$ can be retrieved from the KCDB database [14]. Combined $u_{best}$ standard uncertainties are reported in bold at the bottom of each table, next to $u_{MRA}$ ($k = 1$) values.

Globally, $u_{best}$ uncertainties are evaluated in the same way as $u_{MRA}$. They might be lower than $u_{MRA}$ because they take into account some procedural and technological improvements related to the new calibration facilities introduced in the two laboratories, since the last key comparison or the last CMC review. Table 2 shows that INRIM is able to substantially reduce its uncertainties in the realization of the TPs of e-$H_2$, Ne and $O_2$, with respect to its current CMCs. The improvement is on the one hand the result of the employment of the new automated calibration facility [8], introduced at INRIM after CCT-K2 (see Introduction). Such a new calorimeter provides a well-controlled thermal environment and reduces repeatability uncertainty in measurements. On the other hand, the studies carried out on the isotopic composition effect on the TP temperatures of e-$H_2$ [17] and Ne [18] allow further uncertainty reductions on these two points. LNE-CNAM, instead, has already recently revised its CMCs after the adoption of the new cryostats described in [9] and [14], which explains why its $u_{best}$ and $u_{MRA}$ values are nearly the same. The only exception is at temperatures close to 17 K and 20 K, where $u_{best}$ is roughly half of $u_{MRA}$. The reason is that $u_{MRA}$ values were determined at LNE-CNAM with a liquid-helium cooled cryostat characterized by poor adiabaticity, especially below 24 K, which introduced large thermal uniformity uncertainties. Conversely, $u_{best}$ are the uncertainties obtained in the new cryostat [9], where thermal uniformity has been substantially improved thanks to the use of a cryogen-free cryocooler.

## 5 Conclusion

A comparison has been performed at all the cryogenic fixed points of the range 3.3.1 of the ITS-90 (see Introduction), from 13.8033 K up to 273.16 K, between the temperature laboratories of LNE-CNAM and INRIM, using a CSPRT provided by LNE-CNAM as transfer standard. The (small) temperature differences found are fully compatible with the combined uncertainties of their respective CIPM MRA CMCs. In addition, both NMIs provided their best uncertainty values, which showed good agreement at all the fixed points and are, therefore, worth to be reported in this paper. To further assess such best measurements uncertainties, LNE-CNAM and INRIM are planning a new comparison involving an exchange of fixed point cells, thus adding solidity to the present comparison.

## Acknowledgments

This work was supported by the European Metrology Programme for Innovation and Research (EMPIR) Joint Research Project 18SIB02 Real-K “Realising the redefined kelvin” and as part of an International Technical Cooperation in the framework of the *Joint Research Laboratory for Fluid Metrology Evangelista Torricelli*. The authors want to thank Roberto Gavioso, Daniele Madonna Ripa and Laurent Pitre for their assistance and helpful suggestions and, in particular, Domenico Giraudi, Roberto Dematteis, Giuseppina Lopardo from INRIM and Catherine Martin from LNE-CNAM for the measurements at the triple points of water and mercury.

## References

1. A G Steele, B Fellmuth, D I Head, Y Hermier, K H Kang, P P M Steur, W L Tew, *Metrologia* **39** **(6)** (2002) 551-571, https://iopscience.iop.org/article/10.1088/0026-1394/39/6/6
2. H Preston-Thomas, 1990, *Metrologia* **27** 3, https://www.bipm.org/utils/common/pdf/ITS-90/ITS-90_metrologia.pdf
3. K D Hill, A G Steele, Y A Dedikov, V T Shkraba, *Metrologia* **42** (2005) Techn Suppl 03001, https://iopscience.iop.org/article/10.1088/0026-1394/42/1A/03001
4. K D Hill, A Peruzzi and R Bosma, *Metrologia* **49** (2012) Techn Suppl 03004, https://iopscience.iop.org/article/10.1088/0026-1394/49/1A/03004
5. K D Hill, A Szmyrka-Grzebyk, L Lipinski, Y Hermier, L Pitre and F Sparasci, *Metrologia* **49** (2012) Techn Suppl 03005, https://iopscience.iop.org/article/10.1088/0026-1394/49/1A/03005
6. K D Hill, T Nakano and P Steur, *Metrologia* **52** (2015) Techn Suppl 03003, https://iopscience.iop.org/article/10.1088/0026-1394/52/1A/03003
7. B Fellmuth, L Wolber, D I Head, Y Hermier, K D Hill, T Nakano, F Pavese, A Peruzzi, R L Rusby, V Shkraba, A G Steele, P P M Steur, A Szmyrka-Grzebyk, W L Tew, L Wang, D R White, *Metrologia* **49**(3) (2012) 257-265, https://dx.doi.org/10.1088/0026-1394/49/3/257
8. D Ferri, D Ichim, F Pavese, I Peroni, A Pugliese, F Sparasci, P P M Steur**,** 2$^{nd}$ Seminar on Low-Temperature Thermometry, Wroclaw (Poland), 2003, 102-107
9. F. Sparasci, L. Pitre, G. Rouillé,·J.-P. Thermeau, D. Truong, F. Galet,·Y. Hermier, *"An Adiabatic Calorimeter for the Realization of the ITS-90 in the Cryogenic Range at the LNE-CNAM"*, *International Journal of Thermophysics*, vol. 32, pp 201-214, 2011, Springer Netherlands, https://dx.doi.org/10.1007/s10765-011-0941-y
10. B. Fellmuth, E. Mendez-Lango, T. Nakano, F. Sparasci, "Guide to the Realization of the ITS-90: Cryogenic Fixed Points", https://www.bipm.org/utils/common/pdf/ITS-90/Guide_ITS-90_2_3_Cryogenic_FP_2018.pdf
11. J. V. Pearce, P. P. M. Steur, W. Joung, F. Sparasci, G. Strouse, J. Tamba, M. Kalemci, "Guide to the Realization of the ITS-90: Metal Fixed Points for Contact Thermometry", https://www.bipm.org/utils/common/pdf/ITS-90/Guide_ITS-90_2_4_MetalFixedPoints_2018.pdf
12. F. Sparasci, I. Didialaoui, A. Vergé, Y. Hermier, *"A new calorimeter for the simultaneous calibration of SPRTs and CSPRTs at the triple point of mercury"*, AIP Conference Proceedings 1552, 486 (2013), https://dx.doi.org/10.1063/1.4819589

13. A. Peruzzi, E. Mendez-Lango, J. Zhang, M. Kalemci, "Guide to the Realization of the ITS-90: Triple Point of Water", https://www.bipm.org/utils/common/pdf/ITS-90/Guide_ITS-90_2_2_TPW-2018.pdf
14. KCDB: The BIPM key comparison database, https://www.bipm.org/kcdb/
15. F Pavese, B Fellmuth, D Head, Y Hermier, A Peruzzi, A Szmyrka-Grzebyk, L Zanin, in *Temperature: Its Measurement and Control in Science and Industry*, Vol 7 Part 1, AIP Conf Proc **684** (2003)**,** 161-166, http://dx.doi.org/10.1063/1.1627118
16. F Pavese, D Ferri, I Peroni, A Pugliese, P P M Steur, B Fellmuth, D Head, L Lipinski, A Peruzzi, A Szmyrka-Grzebyk and L Wolber, in *Temperature: Its Measurement and Control in Science and Industry*, Vol 7 Part 1, AIP Conf Proc **684** (2003)**,** 173-178, http://dx.doi.org/10.1063/1.1627120
17. B Fellmuth, L Wolber, Y Hermier, F Pavese, P P M Steur, I Peroni, A Szmyrka-Grzebyk, L Lipinski, W L Tew, T Nakano, H Sakurai, O Tamura, D Head, K D Hill, A G Steele, *Metrologia* **42** (2005) 171-193, http://dx.doi.org/10.1088/0026-1394/42/4/001
18. F Pavese, P P M Steur, Y Hermier, K D Hill, Jin Seog Kim, L Lipinski, K Nagao, T Nakano, A Peruzzi, F Sparasci, A Szmyrka-Grzebyk, O Tamura, W L Tew, S Valkiers, J van Geel, in *Temperature: Its Measurement and Control in Science and Industry*, AIP Conf Proc **1552** (2013) 192-197, http://dx.doi.org/10.1063/1.4821378
19. Technical Annex for the International Temperature Scale of 1990 (ITS-90), https://www.bipm.org/utils/en/pdf/MeP_K_Technical_Annex.pdf
20. https://www.bipm.org/en/committees/cc/cct/publications-cc.html#kelvin-and-temperature-scales
21. P P M Steur, F Pavese, I Peroni, in: *Temperature, its Measurement and Control in Science and Industry*, Vol.7: Proceedings of the Eighth Int. Temperature Symposium, AIP Conf Proc **684,** 2003, 125-130, http://dx.doi.org/10.1063/1.1627112
22. P P M Steur, I Peroni, F Pavese, D Ferri, A Pugliese, 2nd Seminar on Low-Temperature Thermometry, Wroclaw (Poland), 2003, 86-90
23. P P M Steur and D Giraudi, in *Temperature: Its Measurement and Control in Science and Industry*, AIP Conf Proc **1552** (2013) 124-129, http://dx.doi.org/10.1063/1.4819526
24. C Gaiser, B Fellmuth, P Steur, A Szmyrka-Grzebyk, H Manuszkiewicz, L Lipinski, A Peruzzi, R Rusby, D Head, "Euramet.T-K1: Key Comparison from 2.6 K to 24.6 K, using Rhodium-Iron Resistance Thermometers", Tempmeko 2016 (published on KCDB), Metrologia **54**, (2017) Techn Suppl 03002, http://iopscience.iop.org/article/10.1088/0026-1394/54/1A/03002

25. B M Wood and R J Douglas, 1999, *Metrologia* **36** 245, https://iopscience.iop.org/article/10.1088/0026-1394/35/3/7

**Table 1 Calibration results obtained for the CSPRT Rosemount 162D s/n 5435 at INRIM and LNE-CNAM**

| | | INRIM (i) | | | LNE-CNAM (j) | | | | | | MRA | | | best | | |
|---|---|---|---|---|---|---|---|---|---|---|---|---|---|---|---|---|
| Fixed point | $T_{90}$ | $W_i$ | $u_{MRA,i}$ | $u_{best,i}$ | $W_j$ | $u_{MRA,j}$ | $u_{best,j}$ | $W_i$ - $W_j$ | $dT/dW$ | $dT_{ij}$ | $u_p$ | $a$ | $d_{0.95}$ | $u_p$ | $a$ | $d_{0.95}$ |
| | K | | mK | mK | | mK | mK | | K | mK | mK | | mK | mK | | mK |
| TP e-$H_2$ | 13.8033 | 0.00129455 | 0.18 | 0.10 | 0.00129455 | 0.38 | 0.38 | -0.000000003 | 4092 | -0.01 | 0.42 | 2.00 | **1.63** | 0.39 | 2.00 | **1.53** |
| VPP 17 K | 17.0332 | 0.00240893 | 0.78 | 0.78 | 0.00240821 | 2.25 | 1.38 | 0.000000722 | 2188 | 1.58 | 2.38 | 1.95 | **9.33** | 1.58 | 1.90 | **6.54** |
| VPP 20.3 K | 20.2687 | 0.00435417 | 0.78 | 0.78 | 0.00435310 | 2.25 | 1.36 | 0.000001069 | 1329 | 1.42 | 2.38 | 1.96 | **9.24** | 1.57 | 1.92 | **6.38** |
| TP Ne | 24.5561 | 0.00857356 | 0.15 | 0.07 | 0.00857359 | 0.30 | 0.30 | -0.000000025 | 816 | -0.02 | 0.34 | 2.00 | **1.30** | 0.30 | 2.00 | **1.17** |
| TP $O_2$ | 54.3584 | 0.09184703 | 0.13 | 0.06 | 0.09184613 | 0.26 | 0.24 | 0.000000901 | 256 | 0.23 | 0.29 | 1.93 | **1.16** | 0.25 | 1.91 | **1.01** |
| TP Ar | 83.8058 | 0.21597303 | 0.07 | 0.05 | 0.21597235 | 0.22 | 0.21 | 0.000000676 | 230 | 0.16 | 0.23 | 1.95 | **0.91** | 0.21 | 1.94 | **0.85** |
| TP Hg | 234.3156 | 0.84416417 | 0.06 | 0.06 | 0.84416409 | 0.28 | 0.28 | 0.000000081 | 248 | 0.02 | 0.28 | 2.00 | **1.10** | 0.28 | 2.00 | **1.09** |
| TPW | 273.16 | 1 | 0.03 | 0.02 | 1 | 0.20 | 0.09 | 0.000000000 | 251 | 0.00 | 0.20 | 2.00 | **0.78** | 0.10 | 2.00 | **0.39** |

**Table 2 Uncertainty budget of INRIM for CSPRTs calibrations at the fixed points of the ITS-90 range 3.3.1**

| Uncertainty components | Estimate | TP e-$H_2$ | VPP 17 K | VPP 20.3 K | TP Ne | TP $O_2$ | TP Ar | TP Hg | TPW |
|---|---|---|---|---|---|---|---|---|---|
| Fixed point realization (including isotopic effects) | Type B | 0.06 | | | 0.03 | 0.05 | 0.05 | 0.06 | 0.02 |
| Resistance bridge linearity | Type B | 0.01 | | | 0.001 | 0.003 | 0.003 | 0.01 | 0.01 |
| Agreement between reference resistors Rs | Type B | 0.03 | | | 0.01 | | | | |
| Stability of Rs temperature | Type B | 0.03 | | | 0.004 | 0.002 | 0.001 | 0.001 | 0.001 |
| Self-heating | Type B | 0.01 | | | 0.01 | 0.01 | 0.01 | 0.01 | 0.01 |
| Resistance bridge reading | Type A | 0.08 | | | 0.06 | 0.04 | 0.02 | 0.01 | 0.01 |
| SD RIRT 232324 (comparison) | Type B | | 0.75 | 0.75 | | | | | |
| Fit, interpolation and resistance measurements (comparison) | Type A | | 0.20 | 0.20 | | | | | |
| **Combined $u_{best}$/mK ($k$ = 1)** | | **0.10** | **0.78** | **0.78** | **0.07** | **0.06** | **0.05** | **0.06** | **0.02** |
| $u_{MRA}$/mK ($k$ = 1) | | 0.18 | 0.8 | 0.8 | 0.15 | 0.13 | 0.07 | 0.06 | 0.05 |

**Table 3 Uncertainty budget of LNE-CNAM for CSPRTs calibrations at the fixed points of the ITS-90 range 3.3.1**

| Uncertainty components | Estimate | TP e-$H_2$ | VPP 17 K | VPP 20.3 K | TP Ne | TP $O_2$ | TP Ar | TP Hg | TPW |
|---|---|---|---|---|---|---|---|---|---|
| Triple point value | Type B | 0.24 | | | 0.10 | 0.21 | 0.18 | 0.22 | 0.03 |
| Spurious heat flux | Type B | 0.15 | | | 0.08 | 0.05 | 0.05 | 0.13 | 0.03 |
| Hydrostatic pressure effect | Type B | | | | | | | 0.02 | 0.002 |
| Isotopic composition | Type B | 0.20 | | | 0.24 | | | 0.06 | 0.01 |
| Interprestation of the plateau | Type B | 0.04 | | | 0.03 | 0.07 | 0.05 | 0.05 | 0.01 |
| Electrical measurements | Type B | 0.10 | 0.15 | 0.15 | 0.10 | 0.05 | 0.05 | 0.05 | 0.05 |
| Self-heating | Type B | 0.05 | 0.51 | 0.44 | 0.05 | 0.05 | 0.05 | 0.05 | 0.05 |
| Repeatability of the measurements | Type A | 0.09 | 0.02 | 0.14 | 0.02 | 0.04 | 0.05 | 0.03 | 0.03 |
| Reference thermometer calibration (comparison) | Type B | | 1.26 | 1.26 | | | | | |
| Thermal uniformity | Type B | | 0.20 | 0.20 | | | | | |
| **Combined $u_{best}$ / mK ($k$ = 1)** | | **0.38** | **1.38** | **1.36** | **0.30** | **0.24** | **0.21** | **0.28** | **0.09** |
| $u_{MRA}$ / mK ($k$ = 1) | | 0.38 | 2.25 | 2.25 | 0.30 | 0.26 | 0.22 | 0.28 | 0.20 |

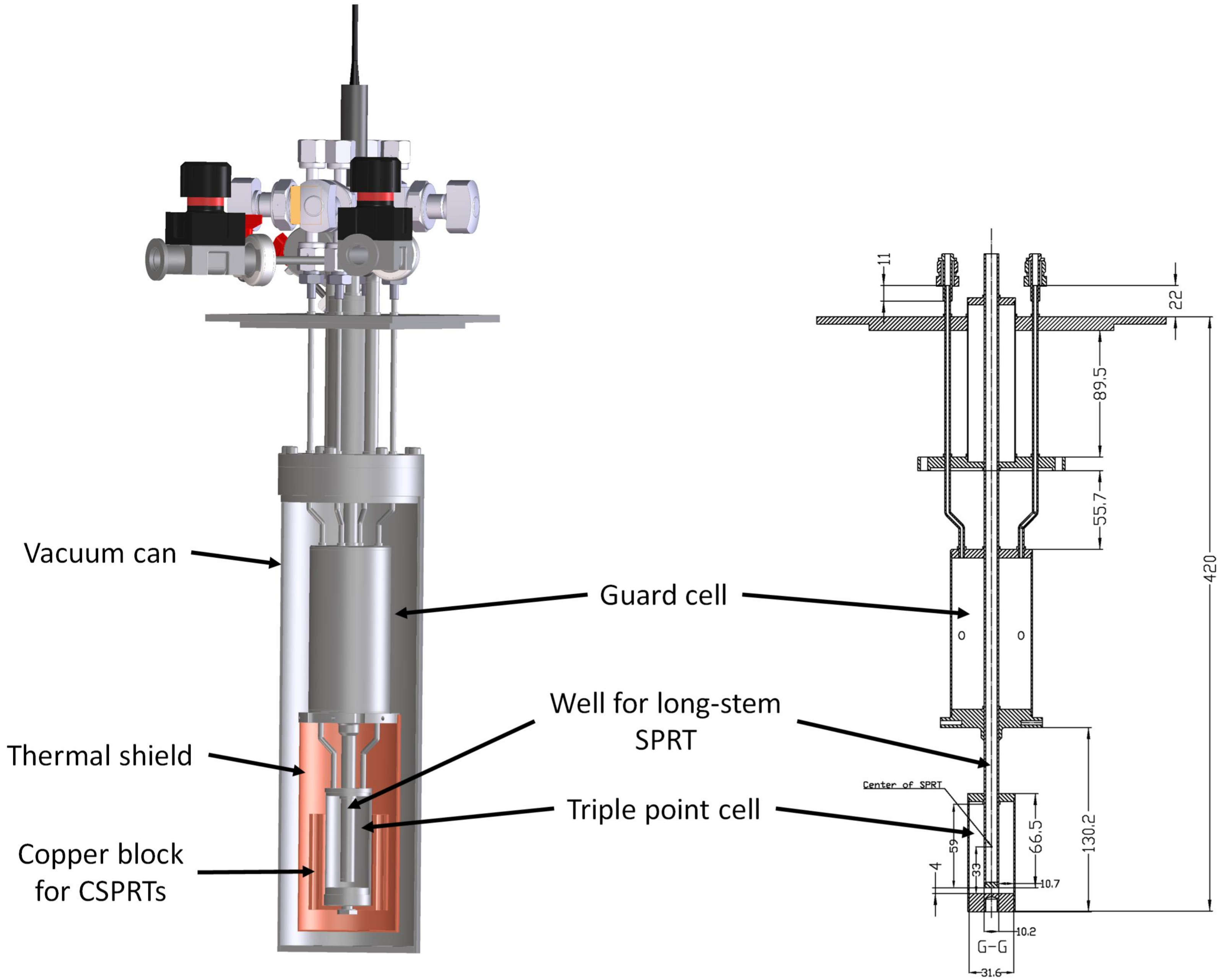

**Figure 1: Layout of the apparatus for the mercury triple point developed at LNE-CNAM**

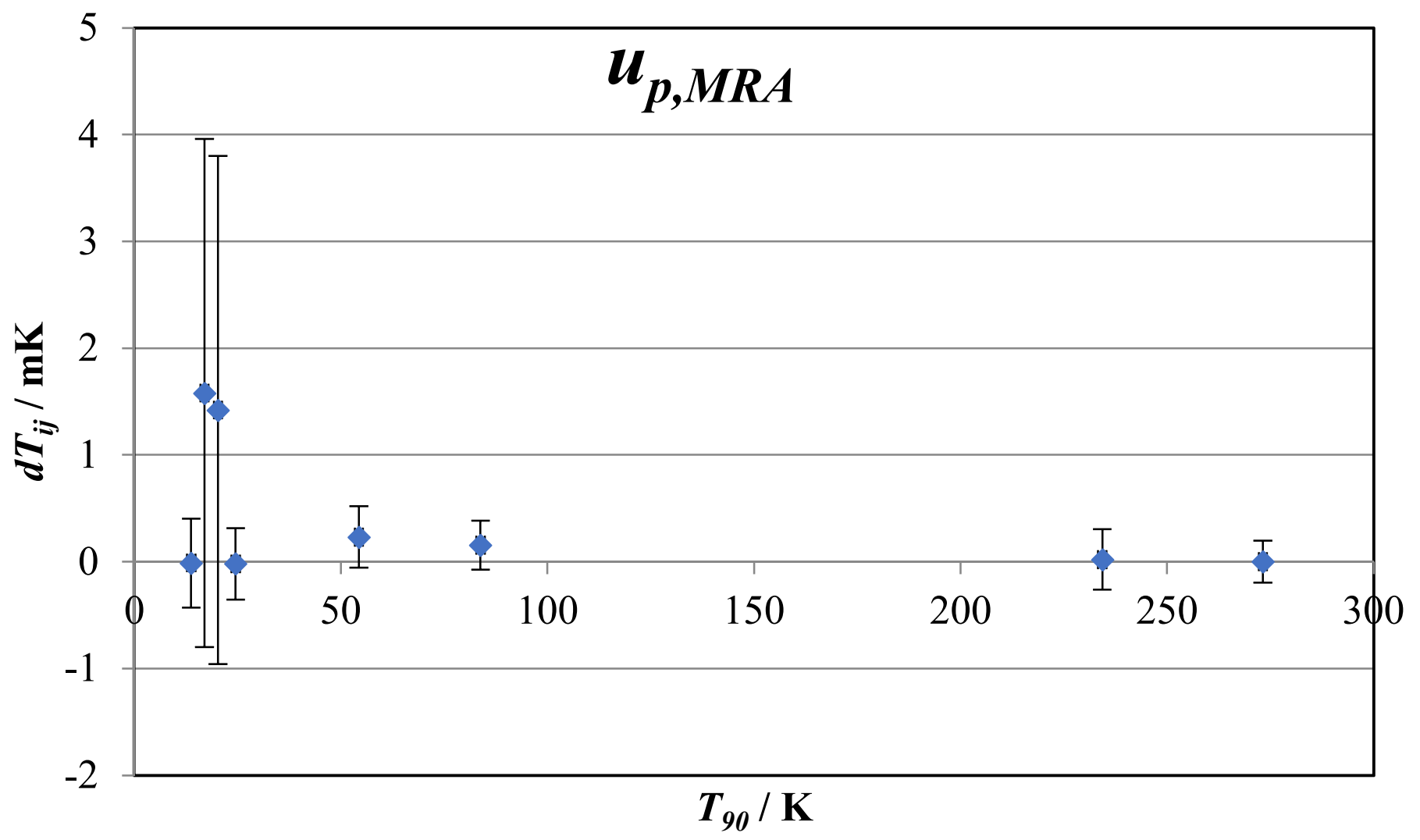

**Figure 2 Difference between LNE-CNAM and INRIM realizations of ITS-90 in its range 3.3.1 ($u_{p,MRA}$ as error bars, $k$ = 1)**

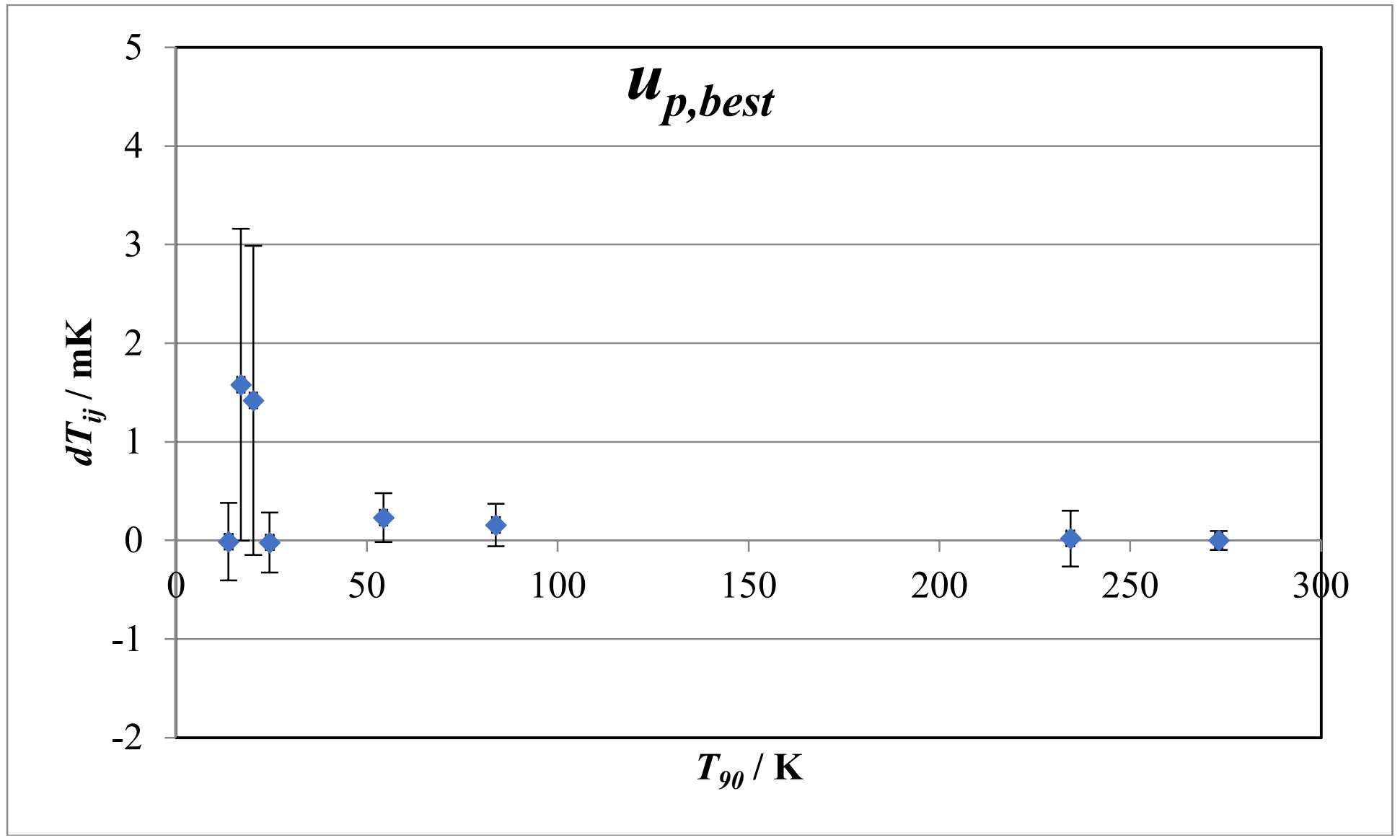

**Figure 3 Difference between LNE-CNAM and INRIM realizations of ITS-90 in its range 3.3.1 ($u_{p,best}$ as error bars, $k$ = 1)**

**Appendix**

As a guide for the reader's eye, the main steps involved in the direct-calibration of CSPRTs are summarized in the following figures as regards the realization of the triple point of equilibrium hydrogen, which are in this case: the ortho-para conversion (Figure A.1), the control preliminary checks (Figure A.2) the realization of the plateau (Figure A.3) and an example of plateau analysis (Figure A.4).

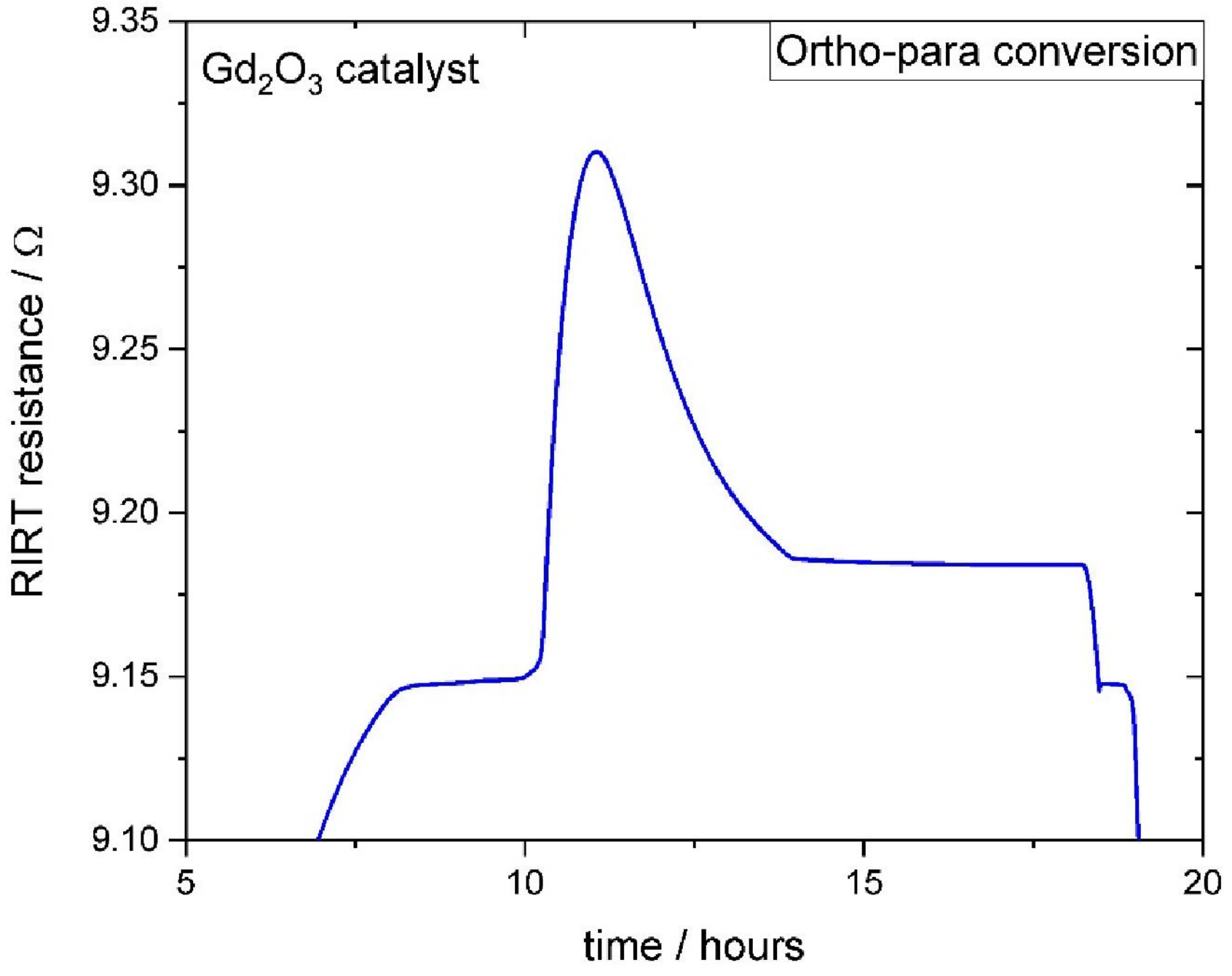

**Figure A.1 Distinctive characteristic, on the resistance trace of a control thermometer, of the conversion of ortho-hydrogen (parallel spins) into para-hydrogen (anti-parallel spins), whose reaching of equilibrium proportions, the horizontal line after the curve, is accelerated by the use of a catalyst inside the cell. The narrow plateaus before and after the curve, at about 9.15 Ohm, form when the triple point temperature is realized**

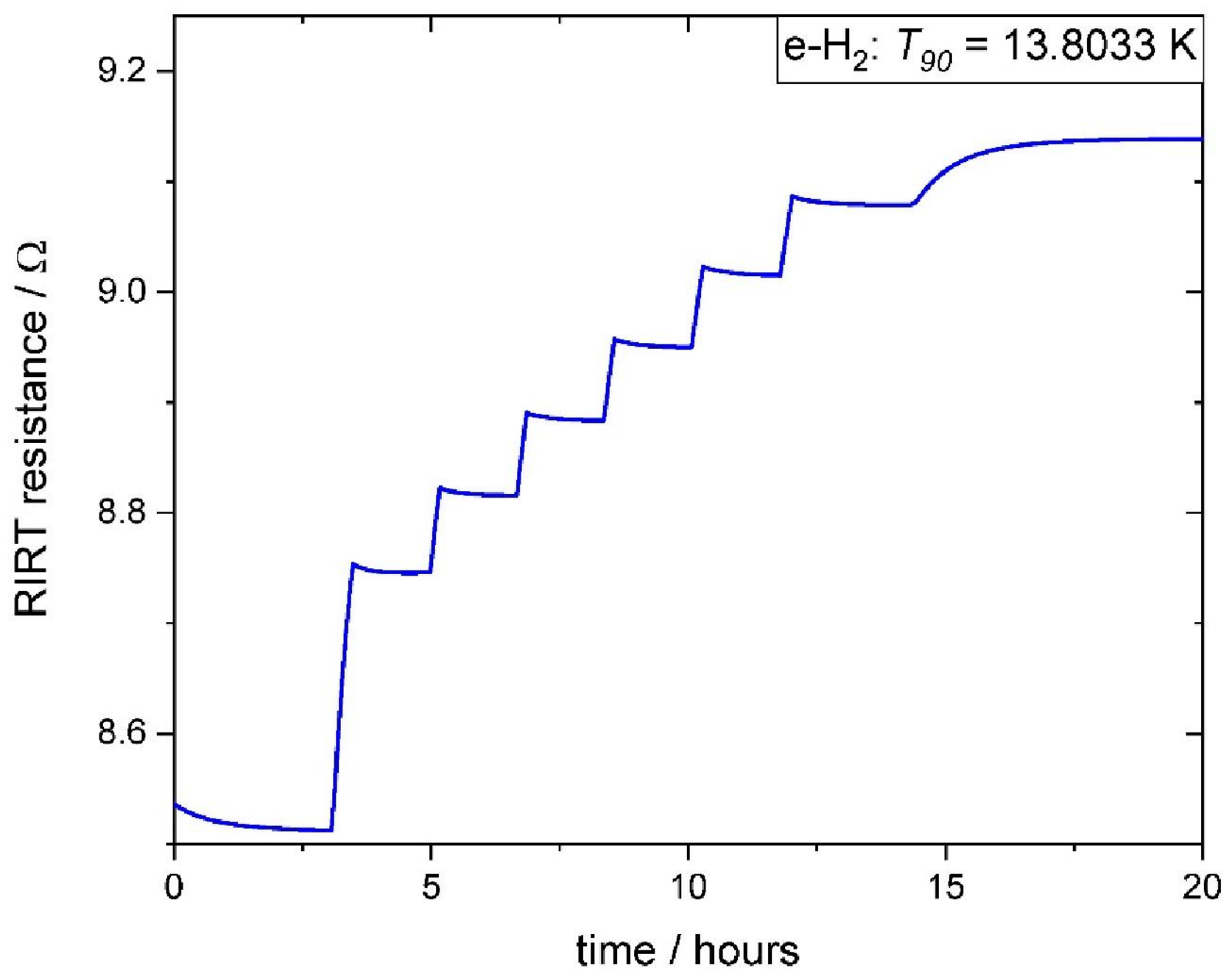

**Figure A.2 Thermal capacity steps and preventive check on the degree of adiabaticity performed prior to the actual realization of the triple point for calibration**

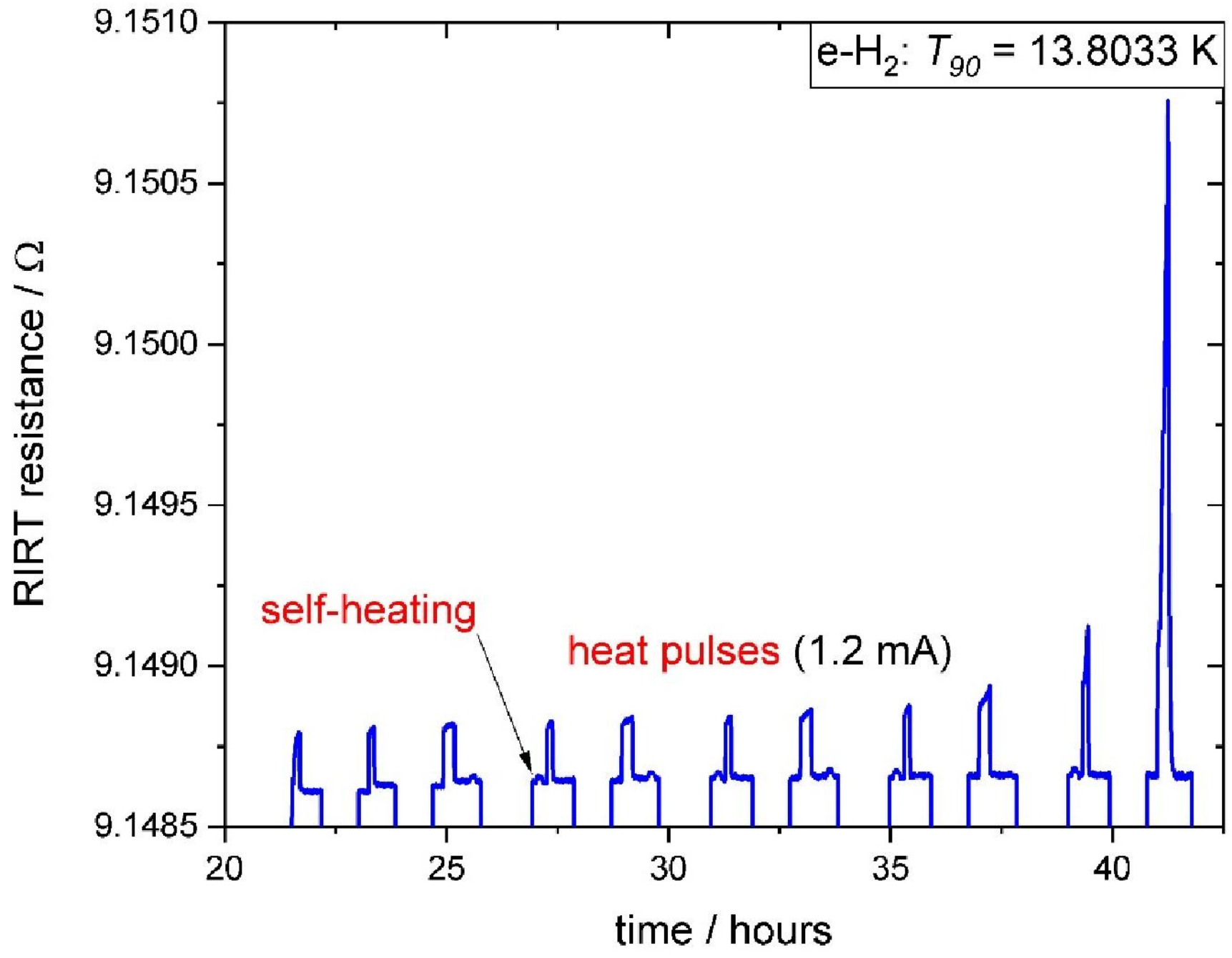

**Figure A.3 A typical plateau at the triple point as it appears during its realization with the calorimetric method. As a consequence of the heat pulses applied, the melted fraction *F* of the substance increases with each heat pulse. The interruptions throughout the plateau correspond to the intervals when the CSPRT to be calibrated is measured. At defined intermediate steps, the self-heating checks are performed**

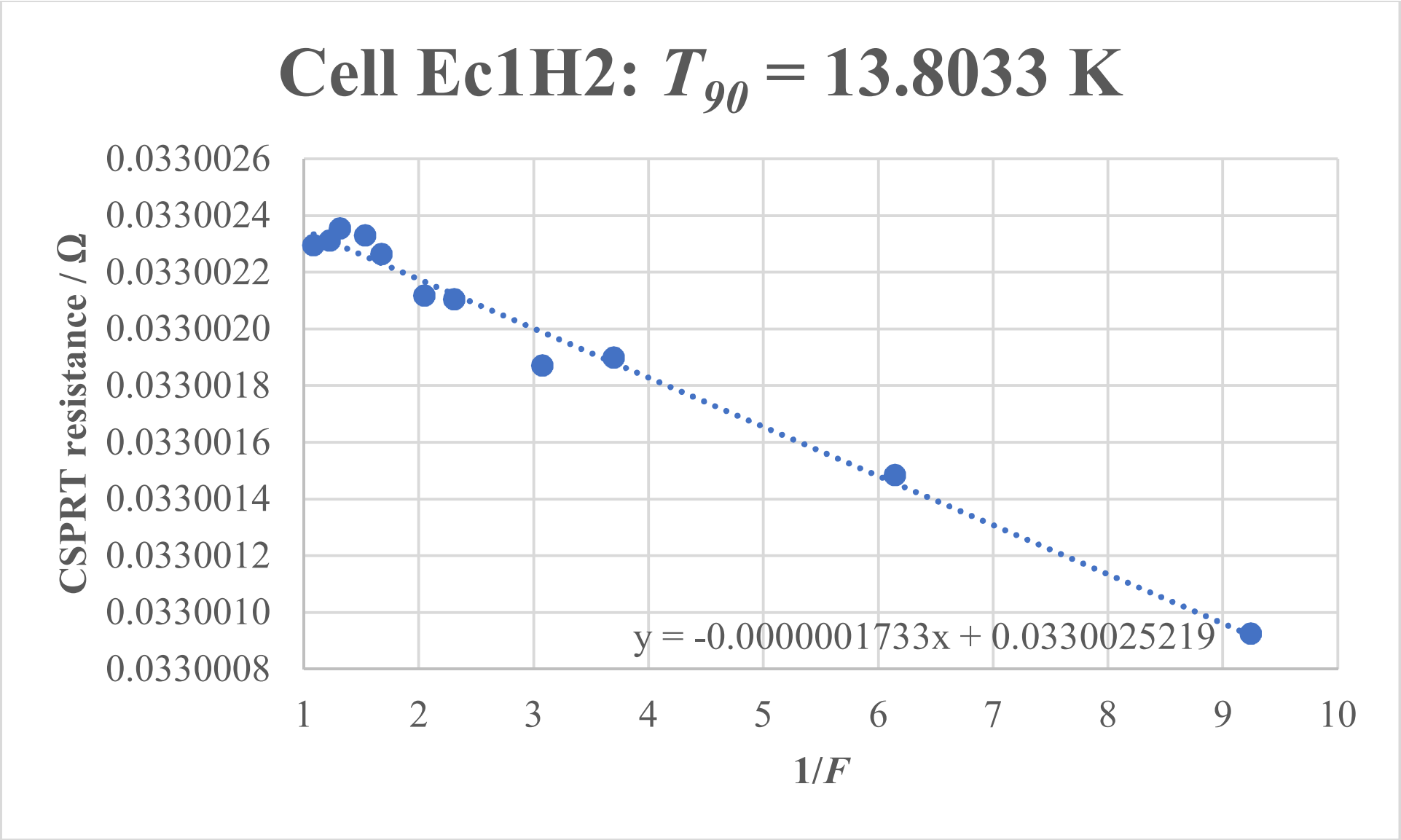

**Figure A.4 The resistance values of the CSPRT measured after each heat pulse, reduced to zero current, are finally plotted against the calculated values of 1/*F* and extrapolated to *F* = 1 (N.B.: the temperature difference between the outer points is about 0.224 mK). After correction for the isotopic effect, the final value of resistance *R* at the triple point is given.**